\documentclass[12pt,preprint]{aastex}

\slugcomment{To appear in ApJ}

\shorttitle{Warm dust in the TPZ of a sun-like Pleiad}
\shortauthors{Rhee et al.}

\begin{document}

\title{Warm dust in the terrestrial planet zone of a sun-like Pleiad: \\
    collisions between planetary embryos? }

\author{Joseph H. Rhee\altaffilmark{1}, Inseok Song\altaffilmark{2}, 
B. Zuckerman\altaffilmark{1,3}}

\altaffiltext{1}{Department of Physics and Astronomy, Box 951547,
University of California, Los Angeles, CA 90095-1562; rhee@astro.ucla.edu,
ben@astro.ucla.edu)}
\altaffiltext{2}{Spitzer Science Center, IPAC/Caltech, Pasadena, CA 91125; 
song@ipac.caltech.edu}
\altaffiltext{3}{NASA Astrobiology Institute}

\begin{abstract}
Only a few solar-type main sequence stars are known to be orbited by warm 
dust particles; the most extreme is the G0 field star BD+20 307 that emits 
$\sim$4\% of its energy at mid-infrared wavelengths.  We report the 
identification of a similarly dusty star HD~23514, an F6-type member of the 
Pleiades cluster.  A strong mid-IR silicate emission feature indicates the 
presence of small warm dust particles, but with the primary flux density 
peak at the non-standard wavelength of $\sim$9\,$\micron$.  The existence 
of so much dust within an AU or so of these stars is not easily accounted 
for given the very brief lifetime in orbit of small particles.  The apparent 
absence of very hot ($\gtrsim$1000 K) dust at both stars suggests the 
possible presence of a planet closer to the stars than the 
dust. The observed frequency of the BD+20 307/HD~23514 phenomenon indicates 
that the mass equivalent of Earth's Moon must be converted, via collisions 
of massive bodies, to tiny dust particles that find their way to the 
terrestrial planet zone during the first few hundred million years of the 
life of many (most?) sun-like stars.  Identification of these two dusty 
systems among youthful nearby solar-type stars suggests that terrestrial 
planet formation is common.
\end{abstract}


\section{Introduction}
The Spitzer Space Telescope is now providing a wealth of new information 
about dusty stars in the Milky Way.  However, because Spitzer is a pointed 
telescope, during its cryogenic lifetime it will examine only a modest 
portion of the sky, about 1\%.  For the very rare, very infrared-bright, 
nearby field star, the less sensitive Infrared Astronomical Satellite 
({\it IRAS}) was actually the better search engine because it was an all-sky 
survey.  Consequently, we have an ongoing program to correlate the 
{\it IRAS} catalog with the Hipparcos \citep{son02,rhe07b} and Tycho 
catalogs \citep{mel08}.  To date we have identified at least 
three nearby stars of age $\gtrsim$100 Myr that emit at least a few percent 
of their energy at infrared wavelengths.  These are: field star BD+20 307 
(\citealt{son05}; A. Weinberger et al 2007, in preparation); a member of a 
field Tycho binary star \citep{mel08}; and the 
Pleiad HD~23514 (= HI~1132) that is the principal focus of the 
present paper.  In addition to these three stars, BP Psc, discovered by 
{\it IRAS} to be very bright at far-IR wavelengths, might also be not young 
and fairly nearby \citep{zuc07}.  In contrast, to the best of our 
knowledge, {\it Spitzer} has not yet discovered any nearby star not in a 
region of recent star formation nearly as infrared luminous as these four.  
By infrared luminosity we mean the fraction of a star's bolometric luminosity 
as seen from Earth that is absorbed and re-radiated by dust particles.  For 
the above-mentioned four stars this fraction is in the range between 2 and 
75\%.

When the remarkable properties of 
BD+20~307 were first appreciated, a statistical analysis of the frequency of 
occurrence of such very dusty stars, based on only one example, might be 
regarded as premature (and no such analysis was attempted by \citealt{son05}).  
However, with the recognition (below) that properties of HD~23514 are quite 
similar to those of BD+20 307, the phenomenon has been transformed from a 
miracle into a statistic.  Consequently, following description of our 
observations, in Section 3 we discuss the occurrence frequency of the very 
dusty phenomenon and what it might imply for the evolution of planetary 
systems in orbit around adolescent-age main sequence stars.

\section{Observations and Results}
The large mid-infrared excess of HD~23514 (Fig. 1) was discovered by IRAS 
only in its 12\,$\micron$ band.  Due to its large beam size, IRAS 
measurements often included many contaminating background sources and 
some IRAS identified IR excess stars were subsequently found to be false 
positives \citep{son02,rhe07b}.  In addition, all previously known dusty 
main sequence stars with excess emission detected at {\it IRAS} 12\,$\micron$, 
also had excess emission in at least one of the longer {\it IRAS} 25, 60, 
or 100\,$\micron$ bands.  This is because the IR detection of cold Kuiper 
Belt analogs has been much more frequent than detection of warm asteroid 
belt analogs.  Thus, significant ambiguity among {\it IRAS} excess candidates 
and lack of known main sequence stars with strong mid-IR excess emission 
perhaps helped to prolong the overlooking of the {\it IRAS} 12\,$\micron$ 
detection of HD~23514.

\citet{spa01} reported a marginal detection of dust excess emission from 
HD~23514 at 60 and 90\,$\micron$ with a pointed observation of the Infrared 
Space Observatory (ISO).  They did mention the {\it IRAS} 12\,$\micron$ 
measurement, but then used only ISO data leading them to an incorrect 
conclusion about the dust properties of this star (T$_{dust} \sim$ 70\,K, 
$L_{dust}/L_{*}$ $\sim$3 $\times 10^{-4}$).  HD~23514 was rediscovered as 
a potential hot dust star from our search of main sequence mid-IR excess 
stars using public Spitzer data (Rhee, Song \& Zuckerman, in preparation).  
Spitzer MIPS 24\,$\micron$ images have a fairly large field of view 
(5\arcmin $\times$ 5\arcmin) and many field stars appear in most MIPS 
images.  HD~23514 was included, serendipitously, in the FEPS [Formation 
and Evolution of Planetary Systems, \citep{mey04}] field of Pleiad HII~1182.  

Follow-up imaging observation of HD~23514 was carried out with the Near 
InfraRed Imager (NIRI) and Mid-IR Imager/Spectrometer (Michelle) at Gemini 
North Telescope.  L$^\prime$ (3.8\,$\micron$) and M$^\prime$ (4.7\,$\micron$) 
images were obtained with NIRI using a four-point dithered pattern.  The 
standard ``beam switching'' mode was used for six mid-IR narrow-band 
images with Michelle by chopping the secondary at 2.7 Hz and nodding 
the telescope every $\sim$30 sec.  For L$^\prime$ and M$^\prime$ 
images, dark frames were first subtracted from raw frames.  After 
sky-subtraction, images were then flat-fielded using a sky frame made by 
median combination of dithered images.  Images at each band were shifted, 
added, and averaged to produce a final image at each wavelength.  For six 
mid-IR images, raw images were sky-subtracted using the sky frame from each 
chop pair.  Subtraction of a nodded pair removed the thermal emission from the 
telescope.  Standard stars, HD 22686 and HD 18884, were observed close in time 
and position to our target and used for absolute flux calibration of 
L$^\prime$ and M$^\prime$ images and of six mid-IR images, respectively.   
Finally aperture photometry was performed on both HD~23514 and the standard 
stars to compute flux density at each band.  We used aperture radii of 
0.96$\arcsec$ and sky annuli of 1.42$\arcsec$ \& 1.97$\arcsec$ for both 
target and the standard star.

These ground-based images at 3.8 $-$ 11.7\,$\micron$ bands show only one 
object at the expected target location (Gemini blind pointing accuracy 
is good to $<1\arcsec$), thus verifying that the dust emission shown 
in Figs. 1 and 2 originates from HD~23514.   Flux measurements at those 
wavelengths confirmed excess emission above the stellar photosphere.  

Near-IR and mid-IR spectra of HD~23514 were obtained using the Near-infrared 
Cross-dispersed Echelle Grating Spectrometer (NIRSPEC, \citealt{mcl98}) 
at Keck II Telescope and Michelle at Gemini North Telescope.  NIRSPEC 
was used in a low resolution mode (R $\sim$ 2000) 
with the 42\arcsec $\times$ 0.570$\arcsec$ slit to obtain a KL (2.8 $-$ 
3.7\,$\micron$) band spectrum of HD~23514.  For N-band spectra, the 
low-resolution spectroscopic mode (R $\sim$ 200) of Michelle was used with 
a 2 pixel-wide (0.402\arcsec) slit.  The N-band filter with a central 
wavelength of 10.5\,$\micron$ was selected to give wavelength coverage 
of 7.7 $-$ 14\,$\micron$.  REDSPEC, an IDL based reduction package for 
NIRSPEC, was used for the reduction of the NIRSPEC KL spectrum.  Dark 
frames were first removed from raw frames.  Using the sky frame of the 
nod pair for KL spectrum and of the chop pair for N-band spectrum, the 
resultant frames were sky-subtracted and flat-fielded.   After 
the spectra of the standard stars (HD~210501 for KL-band and HD~18884 for 
N-band spectra) were divided by Planck curves with each star's effective 
temperature (6400\,K for HD~210501 and 3400\,K for HD~18884), these ratioed 
spectra were then divided into the spectra of HD~23514 to remove telluric
and instrumental signatures.  Wavelength calibration was performed using 
Argon lamp spectrum for the NIRSPEC KL spectrum and using atmospheric 
transition lines from an unchopped raw frame for Michelle N-band spectrum.  
Finally, photometry values at NIRI L$^\prime$ band and Michelle 
8.8\,$\micron$ band were used to flux-calibrate KL and N-band spectra, 
respectively.  Near-IR \& mid-IR photometry of HD~23514 from both 
ground-based and space observations is listed in Table 1.

We note that MSX detected HD 23514 in its A-band (8.28\,$\micron$ isophotal
wavelength) with a catalog flux of 158\,mJy (color uncorrected) 
significantly below our measured fluxes from the ground with Michelle 
narrow bands.  Furthermore, the reported position of the MSX source is 
$\sim$12$\arcsec$ west of HD 23514. To reconcile the discrepancy in flux 
and check the nature of the $\sim$3 sigma positional offset, we downloaded 
a 2$\times$2 degee MSX A-band image of HD 23514.  The image contains about 
two dozen MSX sources.  When we overplotted MSX catalog positions on the 
MSX A-band image, only HD 23514 shows a mysterious shift of $\sim$12$\arcsec$ 
from the obvious source in the image.  All other MSX catalog positions fall 
right on bright sources in the image.  Given the fact that no source other 
than HD 23514 appeared in the 32$\arcsec \times24 \arcsec$ field of Michelle 
images, we attribute the mysterious 12$\arcsec$ offset to an erroneous 
astrometric correction of MSX.  Although the nominal wavelength of the MSX 
A-band is 8.28\,$\micron$, its effective wavelength is dependent on the 
source spectrum (hence the need for color correction).  When a true source 
spectrum is very different from the assumed one ($F_\lambda \sim \lambda^{-1}$)
as in our case, the effective wavelength of the MSX A-band can be significantly
shifted.  To quote an MSX measured flux at its nominal wavelength of 
8.28\,$\micron$, a color correction needs to be applied which can account 
for the apparent discrepancy between our narrow band and MSX catalog fluxes.

In Fig. 1, the slope of the NIRSPEC KL spectrum agrees with the 
dust continuum fit we derive below.  The 8 $-$ 13\,$\micron$ spectrum, 
Figures~1 and 2, reveals warm small dust grains near HD~23514 through a 
prominent emission feature.  For young stars and debris disks, the most 
prominent spectral feature in the N-band is silicate emission. However, our 
N-band HD 23514 spectrum peaking at $\sim$9\,$\micron$ is different from 
almost all other frequently seen silicate features that peak at 9$-$11\,$\micron$ 
due to various combinations of olivine, pyroxene, and other minerals 
(Fig. 2).  Possible carriers of this bizarre 9\,$\micron$ emission 
feature among common minerals in our Solar System and Earth's surface are 
tektosilicates and sulfates. Tektosilicates are a group of light colored 
silicate minerals and this group contains most common minerals (quartz, 
feldspar, etc.) seen on Earth's surface. About 75\% of Earth's crust is 
composed of tektosilicates.  However, explaining the strong 9\,$\micron$ 
feature without accompanying prominent olivine and pyroxene signatures is 
challenging.  For example, it is difficult to imagine an extraordinary amount 
of sulfates at HD 23514 over more commonly appearing minerals unless the 
chemical composition of HD 23514 is very different from solar. The huge quantity 
of dust needed to match the HD 23514 SED can be generated by catastrophic 
collisions among planetary embryos or even a planet-planet collision (see $\S$ 
3).  The latter mimics the postulated Moon-creating collision between the 
young Earth and a Mars size planet \citep{har75,cam76}.  Crustal 
material ejected from such a collision may naturally explain our HD 23514 
N-band spectrum.  However, it is hard to explain how crustal material was 
ejected predominantly over mantle material considering that Earth's crust is 
a thin layer occupying only $<$5\% of volume compared to the mantle.  
Nonetheless, the unusual N-band spectrum of HD 23514 must bear a clue to 
the origin of dust and a wider range mid-IR spectrum is needed for more 
detailed analysis.  At this stage, we likely rule out the case of collisional 
grinding of many asteroids as the source of dust around HD 23514 since olivine 
and pyroxene should be the dominant minerals in such environments.  
We note that few objects out of 111 T Tauri stars investigated by \citet{fur06}, 
who used {\it Spitzer} IRS, show a mid-IR emission feature peaking 
near 9\,$\mu$m (e.g. IRAS 04187+1927 and CZ Tau), as found in 
HD~23514.  

\section{Discussion}
Excess emission peaking at mid-IR wavelengths indicates that dust must be warm 
and close to the central star.   The temperature, as well as the amount of dust 
and its distance from the central star, is constrained by creating a Spectral 
Energy Distribution (SED) assuming that dust exists as an optically thin ring.  
We produced an SED of HD~23514 (Fig. 1) by fitting observed measurements at 
optical and infrared bands with a stellar photosphere model \citep{hau99} and 
a single temperature blackbody of T = 750\,K.  Large blackbody grains in 
thermal equilibrium at 750\,K would be located $\sim$0.25\,AU from HD~23514.  
Even small grains that radiate less efficiently, especially those responsible 
for the mid-IR emission feature, likely lie within a few AU of the central 
star.  We noted in \citet{rhe07a} that stars with warm dust emission, ages 
between 10 and 30\,Myr, and spectral types from G0 to A do not show any 
evidence in their SEDs of the presence also of cold dust.  In such cases, even 
the small dust particles that carry the strong mid-IR emission feature are 
likely located close to the stars.  Likewise, there is no obvious evidence 
for cold dust associated with BD+20 307 (\citealt{son05}; A. Weinberger et 
al.\ 2007, in preparation).  However, this may not 
be the case for HD~23514 because {\it Spitzer} and {\it ISO} points in 
its SED all lie somewhat above the 750\,K dust continuum as shown in Fig.~1, 
thus suggesting the presence of cooler dust 
further from the star.  Still, caution is appropriate in interpretation of 
the {\it Spitzer} 25\,$\micron$ flux density measurement as it may be elevated 
by inclusion of the red wing of a silicate emission feature (e.g., A. 
Weinberger et al.\ 2007, in preparation).  And the 60 and 90\,$\micron$ 
{\it ISO} points, respectively, lie only 2 and 2.3 sigma above the 750\,K 
dust continuum line in Fig. 1.  A Spitzer 70\,$\micron$ flux measurement is 
highly desirable to clarify the presence of cool dust.

A standard method for characterizing the amount of dust orbiting a star is 
through the quantity $\tau$ ($\equiv$ L$_{IR}$/L$_{*}$) where L$_{IR}$ 
is the excess luminosity above the photosphere emitted at infrared 
wavelengths and L$_{*}$ is the bolometric luminosity of the star.  We 
obtained $\tau \sim 2 \times 10^{-2}$ by dividing the infrared excess 
between 2.3\,$\micron$ and 90\,$\micron$ by the stellar bolometric luminosity 
(2.8 L$_{\odot}$); this is $\sim10^{5}$ times greater than that of the Sun's 
current zodiacal cloud ($\tau \sim 10^{-7}$).  HD~23514 thus joins BD+20 307 
as the two Sun-like main sequence stars with by far the largest known 
fractional infrared luminosities (Table 2).  The age of BD+20 307 is at 
least a few hundred million years \citep{son05}.  As a member of the 
Pleiades (HII 1132), the age of HD~23514 is $\sim$100\,Myr.  There can be 
little doubt of cluster membership because in the plane of the sky HD~23514 
is located well inside the cluster\footnote{A picture of the Pleiades Cluster 
with HD 23514 indicated is available from the Gemini Observatory Web site, 
http://www.gemini.edu/index.php?option=content\&task=view\&id=259.}, sharing 
common proper motion, and its radial velocity of 5.9$\pm0.5$\,km\,s$^{-1}$ 
is in good agreement with the velocity of the Pleiades cluster, 
6.0$\pm1.0$\,km\,s$^{-1}$ \citep{liu91}. 

Currently only a handful of stars with ages $>$ 50 Myr show warm excess 
emission (T $\geqq$ 150K), indicative of planetesimals in the terrestrial 
planet zone.  Table 2 lists some parameters of these stars.  BD+20 307 and 
HD~23514 stand out among them with very high dust temperature (T $\geqq$ 
600\,K) and $\tau > 10^{-2}$.  The remaining four warm excess stars have 
cooler dust and $\tau \sim10^{-4}$. 

The SEDs of HD~23514 and BD+20 307 exhibit excess near-IR emission beginning 
at wavelengths $\sim$4\,$\micron$.  Given the potential importance of stellar 
wind drag on the dust particles (see below), the absence of really hot dust 
($\gtrsim$1000\,K), suggests the possible presence of a ``sweeper planet'' closer 
to the stars than the dust.  Such a situation pertains at HD 69830 where, 
also, no very hot dust is seen and where Neptune-mass planets interior to 
the dust disk are known to exist from precision radial velocity measurements 
\citep{lov06}.  

Using a flat disk model, \citet{jur03} and \citet{jur07} have successfully 
reproduced the IR emission from flat, geometrically thin, dust disks orbiting 
some white dwarfs.  This geometry implies the absence of significant 
gravitational perturbations by objects with substantial mass located in the 
vicinity of the dust.   In contrast, we find that a flat disk of dust 
particles generates insufficient mid-IR flux to match the SEDs of HD~23514 
and BD+20 307.  Thus, the dust orbiting these stars is Òpuffed upÓ in the 
vertical direction, perhaps as a result of the gravitational field of the 
above mentioned sweeper planet, or the gravity of planetary embryos as 
discussed below, or both.

Given the ages of HD~23514 and BD+20 307, it is natural to ask whether their 
huge warm dust burdens were generated by events analogous to those that 
occurred during the ``Late Heavy Bombardment'' (LHB) in our solar system.  
One current model \citep{gom05} attributes the LHB to a rapid migration of 
the giant planets that destabilized the orbits of objects in the Kuiper Belt 
and the main asteroid belt hundreds of millions of years after the formation 
of the Sun.  \citet{wya07} have proposed a similar model for most of the 
stars listed in our Table 2, with the difference that they strongly favor an 
origin of the parent bodies in a region more analogous to the cold Kuiper Belt 
than the asteroid belt.  We note, however, there is no evidence for cold dust 
at BD+20 307, $\zeta$~Lep, and HD~69830, cold dust that might reasonably be 
expected at stars with so much warm dust should all parent bodies originate 
in distant cold regions.  The situation is more ambiguous for a Table 2 star 
like $\eta$ Crv with clear evidence for substantial amounts of cold dust 
(see Fig. 3).  Furthermore, \citet{cuk06} argue that LHB was a localized 
Earth-Moon system activity rather than a global, inner solar system, event.  
Thus, because the cause(s) of the LHB remain unsettled, we do not further pursue 
a relationship between LHB and the high $\tau$ warm excess phenomenon.  

Initially, when there was only one known main sequence star with very large 
$\tau$ ($> 10^{-2}$), the BD+20 307 phenomenon might have been regarded as 
a ``miracle'', so that a statistical analysis of the occurrence rate would 
have been of questionable value.  Now, however, with HD~23514, the frequency 
of occurrence of such extraordinarily dusty stars can be treated statistically 
more reliably.  BD+20 307 is a field star with estimated age similar to that 
of the Ursa Majoris moving group \citep{son05}, whose age is probably about 
400\,Myr \citep{zuc06}.  {\it IRAS} was sufficiently sensitive to detect main 
sequence G-type stars with $\tau > 10^{-2}$ out to $\sim$150 pc.  There are 
$\sim$18400 Hipparcos dwarfs with spectral types between F4 and K0 out to 
130 pc, the distance to the Pleiades.   But most of these are old stars.  If 
the star formation rate was approximately uniform during the past 5 Gyr years, 
then there are about $\sim$1800 Hipparcos dwarfs of age $\sim$500 Myr out to 
130 pc.  Then a Hipparcos dwarf with $\tau > 10^{-2}$ is found among 
solar-type field stars about one time out of 1500 (after dropping $\sim$300 
Hipparcos-measured members of nearby stellar clusters).  

In the solar vicinity several stellar clusters have solar-type members.  We 
select four rich nearby clusters with ages 70$-$700 Myr: Hyades, Pleiades, 
$\alpha$ Persei, and Praesepe.  In these stellar clusters, there are about 
$\sim$400 dwarfs stars with spectral types between F4 \& K0 
\citep{deb01,sta07,lod05,ada02}.  Among these only HD~23514 is identified 
with $\tau$ $> 10^{-2}$.  Therefore the occurrence rate of a dwarf with 
$\tau > 10^{-2}$ in the nearby stellar clusters is, at most, about one out 
of 400.  Combining this result with that for BD+20 307, indicates that the 
very high $\tau$, warm dust, phenomenon manifests itself in about one 
adolescent star (age a few 100 Myr) in 1000.  If all F4-K0 stars display 
this phenomenon as adolescents, then the lifetime of the phenomenon at a 
typical solar-like star is a few 100,000 years.  

To interpret our observations, we consider a model of colliding planetary 
embryos.  In a series of papers, Agnor, Asphaug, and colleagues 
\citep{agn99,agn04,asp06} considered the collisions of large bodies in the 
late stages of the formation of planets in the terrestrial planet zone.  
Based on their models and those they attribute to earlier researchers (e.g., 
G.W. Wetherill), we may draw the following conclusions.  The process of 
terrestrial planet formation involves the formation of a minimum of many 
hundreds of planetary embryos of dimensions $\gtrsim$1000 kilometers.  These 
collide and either they coalesce or, oftentimes, the smaller embryo fragments 
into smaller objects along with the ejection of ``copious debris''.  While 
the mass spectrum of the fragments is not well constrained, no large 
monoliths survive following disruption of large solid bodies.  Rather a 
typical large fragment size might be $\sim$100\,m.  Collisions of planetary 
embryos continue for as long as a few 100\,Myr, i.e., to the ages of HD~23514 
and BD+20 307; in the following discussion, we assume these to be 100 and 
400\,Myr, respectively.  Additional discussion of catastrophic fragmentation 
of planetary system bodies of moderate size may be found in \citet{fuj80} and 
\citet{hou90}.

For these assumed ages and an occurrence rate of one in 1000 stars, we will 
use a lifetime of 250,000 years for the HD~23514/BD+20 307 phenomenon at a 
typical adolescent-age solar-type star.  Small particles now in orbit around 
these two stars will be lost in a much shorter time span and must be 
replenished many times over.  One possible loss mechanism is a collisional 
cascade that breaks particles down in size until, when their radii become 
as small as a few tenths of a micron, they become subject to radiation 
pressure blowout.  Other loss mechanisms are Poynting-Robertson (PR) and 
stellar wind drag.  As mentioned above, we assume that the initial mass 
spectrum is a result of the collision of two planetary embryos, but the 
spectrum of the collision fragments is not well characterized.  Therefore, 
we assume that collisions are sufficiently frequent to establish an 
approximately equilibrium size distribution:
\begin{equation}
N(a) da = N_o a^{-3.5} da
\end{equation}
where N(a)  is the number of particles per $cm^{3}$ with radii between $a$ 
and $a+da$ (see, e.g., \citealt{doh69,wil94,che01}).  The smallest 
particle radius in this distribution may be set by radiation pressure blow-out; 
for the mass (1.35\,M$_{\odot}$) and luminosity (2.8\,L$_{\odot}$) we 
estimate for HD 23514, this radius is $\sim$0.5 microns.

With this size distribution most of the mass (M) is carried by the largest 
particles while most of the surface area ($\tau$) is due to small particles 
with radii not much larger than the submicron-size blowout radius.  
Specifically, 
\begin{equation}
M  \sim \int \frac{4\pi}{3} a^3 a^{-3.5} da  \sim a^{1/2}
\end{equation}
\begin{equation}
\tau(a) \sim \int N(a) \pi a^2 da  \sim \int \pi a^2 a^{-3.5} da \sim a^{-1/2}
\end{equation}
and          
\begin{equation}
t_c = P/\tau \sim a^{1/2}
\end{equation}
where $t_c$ is the collisional lifetime and P is the orbital period.  
Thus, the smallest particles collide the 
fastest, for HD~23514 in about 50 years at 1\,AU.  Larger objects take longer 
to collide destructively.  They will then be broken down into smaller 
fragments such that after a collision of two roughly equal mass objects the 
largest left-over fragment has a radius about 1/2 that of a collider (S. Kenyon 
2007, private comm.).  As more mass is carried by the larger colliders, with 
the above N(a), the ratio of mass to collision time is independent of $a$, 
and there will thus be a supply of material approximately constant with time 
as the largest objects are eventually whittled down to micron size dust.  In 
the Agnor/Asphaug picture outlined above, the largest initial fragments of 
a collision of planetary embryos might have $a \sim$ 100\,m.  Thus, if the 
lifetime of $a$ = 1\,$\micron$ particles is $\sim$50 years due either to 
collisions or to stellar wind drag (see below), then the lifetime of 100 m 
fragments will be $\sim$500,000 years.  Thus, catastrophic disruption of 
large planetary embryos (see mass estimate below) can supply material for a 
time equal to the 250,000-year event lifetime indicated by our observations.

In addition to collisions, PR and stellar wind drag are loss mechanisms for 
small dust particles.  The time scale for PR drag at HD~23514 may be 
evaluated using, for example, Equ. 5 in \citet{che01}.  We assume L$_{*}$ = 
2.8 L$_{\odot}$ (corresponding to a distance of 130 pc to HD~23514), a dust 
particle orbital semi-major axis of one AU, and a radius and density of a 
typical individual grain to be 1\,$\micron$ and 2.5 g/cm$^3$, respectively.  
Then the lifetime against PR drag is $\sim$1000 years.  For the present day 
Sun, PR drag acts about 3 times more rapidly than drag due to the solar wind 
\citep{pla05}.  However, \citet{woo02} estimated that the winds of solar type 
stars decline as time to the 2.00 $\pm$ 0.52 power beginning at ages 
$\sim$10\% of the current age of the Sun.  More recent observation and 
analysis somewhat cloud quantitative representation of the wind strength as 
a function of time \citep{woo05,woo06}.  Based on these references, we 
assume that, for adolescent-age stars like HD~23514 and BD+20 307, wind drag 
will dominate PR drag by a factor $10-30$.  Then the lifetime of 
the small particles considered above against wind drag will be only 
$\sim$50 years, i.e., comparable to the collision times.  Because PR and 
wind drag times are proportional to $a$, while the collision time is 
proportional to $a^{1/2}$, the orbital lifetimes of large particles (rocks) 
are determined by collisions.

To determine how rapidly mass is lost due to either stellar wind drag or 
collisions we estimate the minimum dust mass ($M_{min}$) needed to intercept 
2\% of the light emitted by HD~23514.    This is
\begin{equation}
M_{min} = 16 \pi \tau a \rho R^2 / 3
\end{equation}
Where $\tau$ = 0.02, $a$ = 1\,$\micron$, $\rho$ = 2.5\,g/$cm^3$, and R = 1\,AU.
Then $M_{min} \sim 2\times 10^{22}$ g, with a corresponding mass loss 
rate, $\dot{M} \sim 10^{13}$ g/s.  For BD+20 307, $\tau$ is twice as 
large, so $\dot{M}$ is $\sim$2 $\times 10^{13}$ g/s.  In 250,000 years, 
the total mass lost per star will be $\sim$10$^{26}$ g$-$the mass of 
Earth's moon. Or, for the above assumed average 
density, an object with radius $\sim$2000 km.  Of course, this mass need not 
all be produced in one single catastrophic collision, but might rather be a 
consequence of multiple collisions of smaller planetary embryos spaced over 
hundreds of millions of years.

\section{Conclusions}
We show that substantial quantities of warm, small, dust particles orbit 
HD~23514, a solar-type member of the 100\,Myr old Pleiades cluster.  A 
similar phenomenon was previously reported \citep{son05} for the somewhat 
older, solar-type, field star BD+20 307.  Models for catastrophic collisions 
of planetary embryos orbiting $\sim$100 Myr old stars (e.g., 
\citealt{ken05,asp06}) will naturally produce such warm dusty disks.  Our 
data are consistent with these model predictions, provided that such 
catastrophic events followed by a subsequent collisional cascade convert of 
order the mass equivalent of Earth's Moon to tiny dust particles during 
the early lifetime of many (perhaps most) Sun-like stars.  For example, 
in the case of our solar system, by itself the collision that is postulated 
to have generated the Moon likely would have sent a comparable mass of 
debris into interplanetary orbits.   Infrared data for stars such as 
HD~23514 and BD+20 307 are consistent with and may well validate the 
standard picture of violent formation of terrestrial-like planets in the 
early years of planetary systems.

\acknowledgments
We thank A. Weinberger \& E. E. Becklin for permission to use some 
Michelle data obtained during a joint Keck time-exchange observing run, 
and E. Rice for assistance in obtaining the NIRSPEC spectrum.  
We also thank S. Kenyon and C. Lisse each for a helpful discussion and 
are grateful to C. H. Chen for providing a {\it Spitzer} IRS spectrum 
of $\eta$ Corvi.  We appreciate the constructive comments of the referee.  
Based on observations (GN-2006A-Q-39 and GN-2006B-DD-5) 
obtained at the Gemini Observatory, which is operated by the Association 
of Universities for Research in Astronomy, Inc., under a cooperative agreement
with the NSF on behalf of the Gemini partnership: the National Science 
Foundation (United States), the Science and Technology Facilities Council 
(United Kingdom), the National Research Council (Canada), CONICYT (Chile), 
the Australian Research Council (Australia), CNPq (Brazil) and SECYT 
(Argentina).  This research was supported in part 
by NASA grants to UCLA.  This research has made use of the VizieR and of 
data products from the Two Micron All Sky Survey.


\clearpage

\begin{deluxetable}{lcccc}
\tablecaption{HD 23514 near-IR \& mid-IR Photometry}
\tablenum{1}
\tablewidth{0pt}
\tablecolumns{4}
\tablehead{
\colhead{Filter}        &\colhead{Central Wavelegth} 
&\colhead{Flux Density} &\colhead{Uncertainty} 
&\colhead{}             \\
\colhead{}              &\colhead{($\mu$m)} 
&\colhead{(mJy)}         &\colhead{(mJy)} 
&\colhead{Instrument}
}
\startdata
B               &0.44      &466\tablenotemark{a}       &110\tablenotemark{b}      & TYCHO-2\\
V               &0.55      &600\tablenotemark{a}       &86\tablenotemark{b}       & TYCHO-2\\
J               &1.25      &670       &15       & 2MASS\\
H               &1.65      &516       &11       & 2MASS\\
K$_{s}$         &2.20      &369       &9        & 2MASS\\
L$^\prime$      &3.78      &196       &10  	& Gemini NIRI\\
M$^\prime$      &4.68      &175       &7   	& Gemini NIRI\\
Si-1            &7.7       &197       &10	& Gemini Michelle\\
Si-2            &8.8       &234       &12	& Gemini Michelle\\
Si-3            &9.7       &209       &10	& Gemini Michelle\\
Si-4            &10.3      &140       &7	& Gemini Michelle\\
Si-5            &11.6      &104       &5	& Gemini Michelle\\
Si-6            &12.5      &95        &5	& Gemini Michelle\\   
12$\mu$m        &11.5      &184       &26	& IRAS\\   
24$\mu$m        &24.0      &66.5      &2.7	& MIPS\\   
60$\mu$m        &60.0      &28        &10	& ISO\\   
90$\mu$m        &90.0      &26        &10	& ISO\\   
\enddata
\tablenotetext{a}{The standard $B$ and $V$ magnitudes were obtained by converting 
Tycho~$B$ and $V$ magnitudes using Table 2 in \citet{bes00}.}
\tablenotetext{b}{$B$- \& $V$- flux density uncertainties were 
computed assuming 0.200 for their magnitude uncertainties in order to compensate 
for some missing opacity species in the model spectrum (see \citealt{rhe07b}.)}
\end{deluxetable}

\clearpage

\begin{deluxetable}{ccccccc}
\tablecolumns{7}
\tablenum{2}
\tablecaption{Main sequence stars with debris systems in the terrestrial 
planetary zone \label{tbl-2}}
\tabletypesize{\small}
\tablewidth{0pc}
\tablehead{
\colhead{}               &\colhead{}           &
\colhead{Dust}           &\colhead{}           &
\colhead{}               &\colhead{}           &
\colhead{}               \\
\colhead{}               &\colhead{Spectral}   &
\colhead{Temperature}    &\colhead{$\tau$}     &
\colhead{Age}            &\colhead{Cold}       &
\colhead{}               \\
\colhead{Object}         &\colhead{Type}       &
\colhead{(K)}            &\colhead{($\times 10^{-4}$)}   &
\colhead{(Myr)}          &\colhead{dust}       &
\colhead{References\tablenotemark{a}}     \\
}
\startdata
HD 23514\tablenotemark{b}       & F6V  & 750       & 200  & 100  & Maybe & 1  \\
BD+20 307\tablenotemark{c}      & G0V  & 650       & 400  & 400  & no    & 2  \\
$\zeta$ Lep    & A3   & 190       & 0.65 & 300  & no    &  3,4 \\
HD 72905\tablenotemark{d}       & G1.5 & ?         & 1    & 400  & yes   & 5,6  \\
$\eta$ Corvi   & F2V  & 180 \& 30 & 5    & 600  & yes   &  4,6,7 \\
HD 69830\tablenotemark{d}       & K0V  & ?         & 2    & 2000 & no    & 6,8  \\
\enddata
\tablecomments{We define the terrestrial planet zone (TPZ) to be the 
region where dust particles that radiate like blackbodies will attain 
a temperature of at least 150\,K (see \citealt{rhe07a}).}
\tablenotetext{a}{
1. This paper
2. \citet{son05}
3. \citet{che01}
4. \citet{che06}
5. \citet{bei06}
6. \citet{wya07}
7. \citet{wya05}
8. \citet{bei05}}
\tablenotetext{b}{Our fit to the SED in Fig. 1 implies a photospheric 
temperature of 6400\,K and a stellar radius = 1.28\,$R_{\odot}$ for an 
assumed distance to HD 23514 of 130 pc.  However, based on pre-main 
sequence evolution models of Baraffe et al.(1998,2002), a $\sim$100\,Myr 
old 6400\,K star has radius 1.38\,$R_{\odot}$ and mass 1.35\,$M_{\odot}$.  
If HD~23514 is a single star, then the discrepancy between the two radii 
would be eliminated if the actual distance to HD~23514 is $\sim$140\,pc.  
The spectral type of HD~23514 is listed as F5 in \citet{gra01} while 
\citet{cox00} gives F7 for a main sequence star with T = 6400 K.  We 
adopt F6 in this paper.}
\tablenotetext{c}{T = 6000\,K, R = 1.25\,$R_{\odot}$ \citep{son05}}
\tablenotetext{d}{Following \citet{wya07}, we list HD~69830 \& 
HD~72905 as potentially having dust particles in the TPZ.  \citet{bei05} 
and \citet{lis07} fit a complex model to the mid-IR spectrum of 
HD~69830 and derive an underlying dust continuum temperature of 400\,K.  
However, given the number of free parameters included, and not included 
(e.g. particle shape), in these models we regard the dust temperature 
as not well constrained.}
\end{deluxetable}

\clearpage

\begin{figure}
\plotone{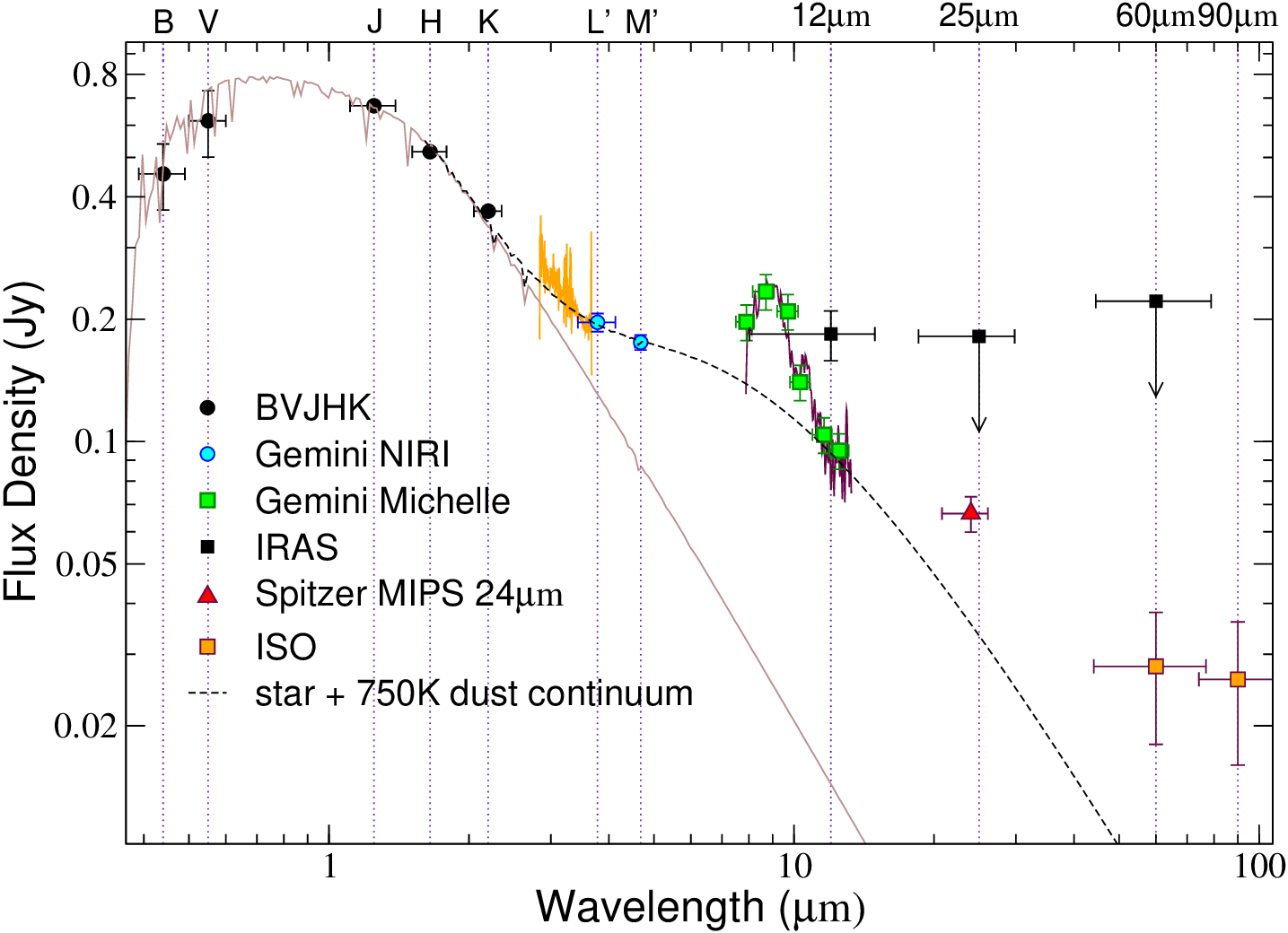}
\caption{Spectral Energy Distribution of HD 23514.  NIRSPEC spectrum is 
shown in yellow (wavelengths 2.8 $-$ 3.7\,$\micron$), while the Michelle 
spectrum is in maroon (7.8 $-$ 13.3\,$\micron$).   The derived stellar 
parameters for HD 23514 are given in a note to Table 1.  For each 
measurement the horizontal bars indicate the passband of the filter used 
and the vertical bars depict 1\,$\sigma$ flux uncertainties.}
\end{figure}

\clearpage

\begin{figure}
\plotone{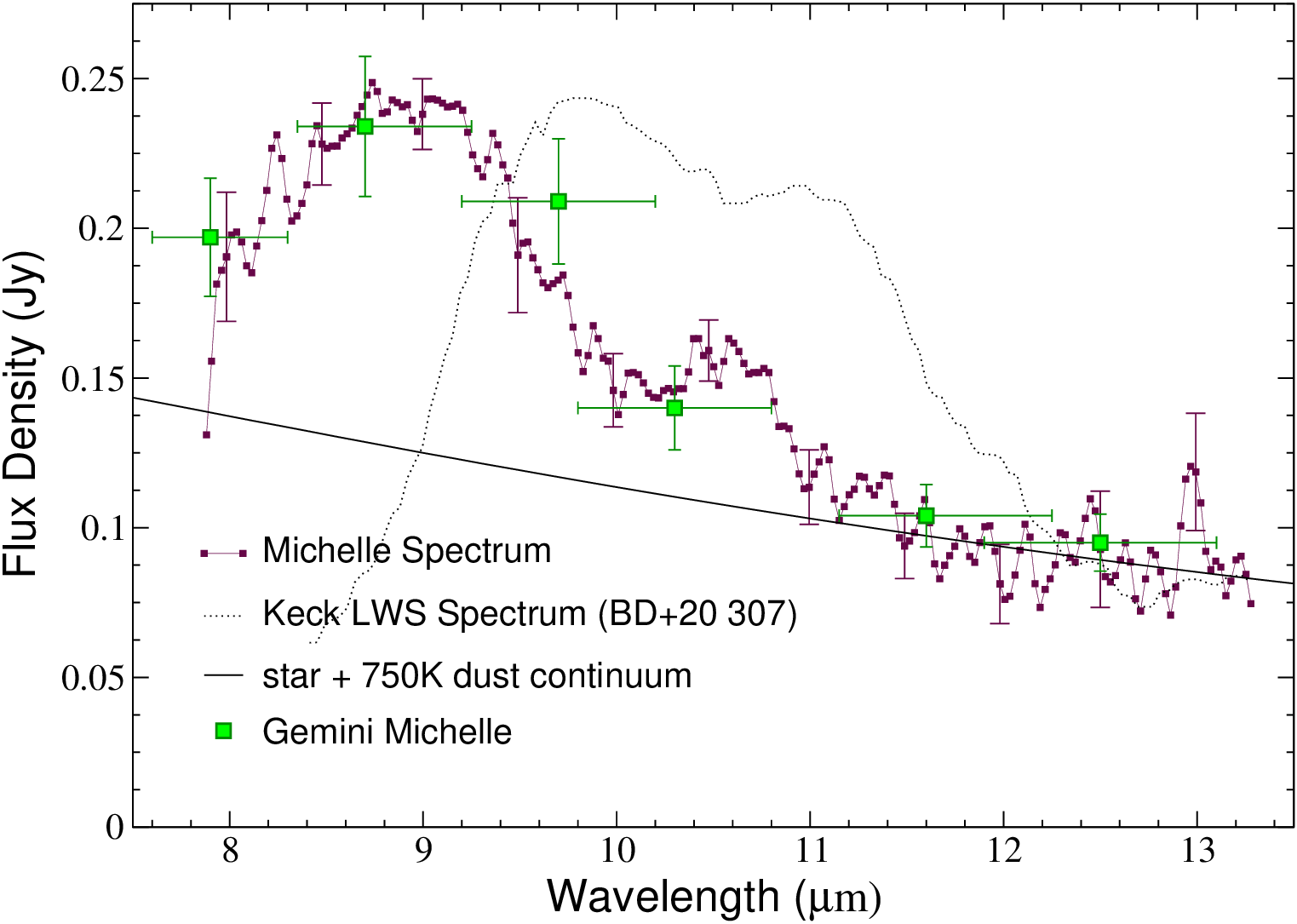}
\caption{Michelle mid-IR spectrum of HD 23514.  A 2\,$\arcsec$ wide sky 
region $\sim$3\,$\arcsec$ from the HD 23514 spectrum was used to compute 
the uncertainty at each wavelength.  Representatve 1\,$\sigma$ uncertainties 
of the spectrum (vertical bars) are given at each half micron.  For 
comparison, the mid-IR, dotted-line, spectrum of BD+20 307 is reproduced 
from Fig. 1 in \citet{son05}.  The spectrum of HD 23514 has almost twice 
the spectral resolution as the Keck LWS spectrum of BD+20 307.  For each 
photometry measurement (green square) the horizontal bars indicate the 
passband of the filter used and the vertical bars depict 1\,$\sigma$ 
uncertainties.  As noted in $\S$ 2, the shape and location of the BD+20~307 
feature is similar to that unsually seen in pre-main sequence stars and 
solar system comets.
}
\end{figure}

\clearpage

\begin{figure}
\plotone{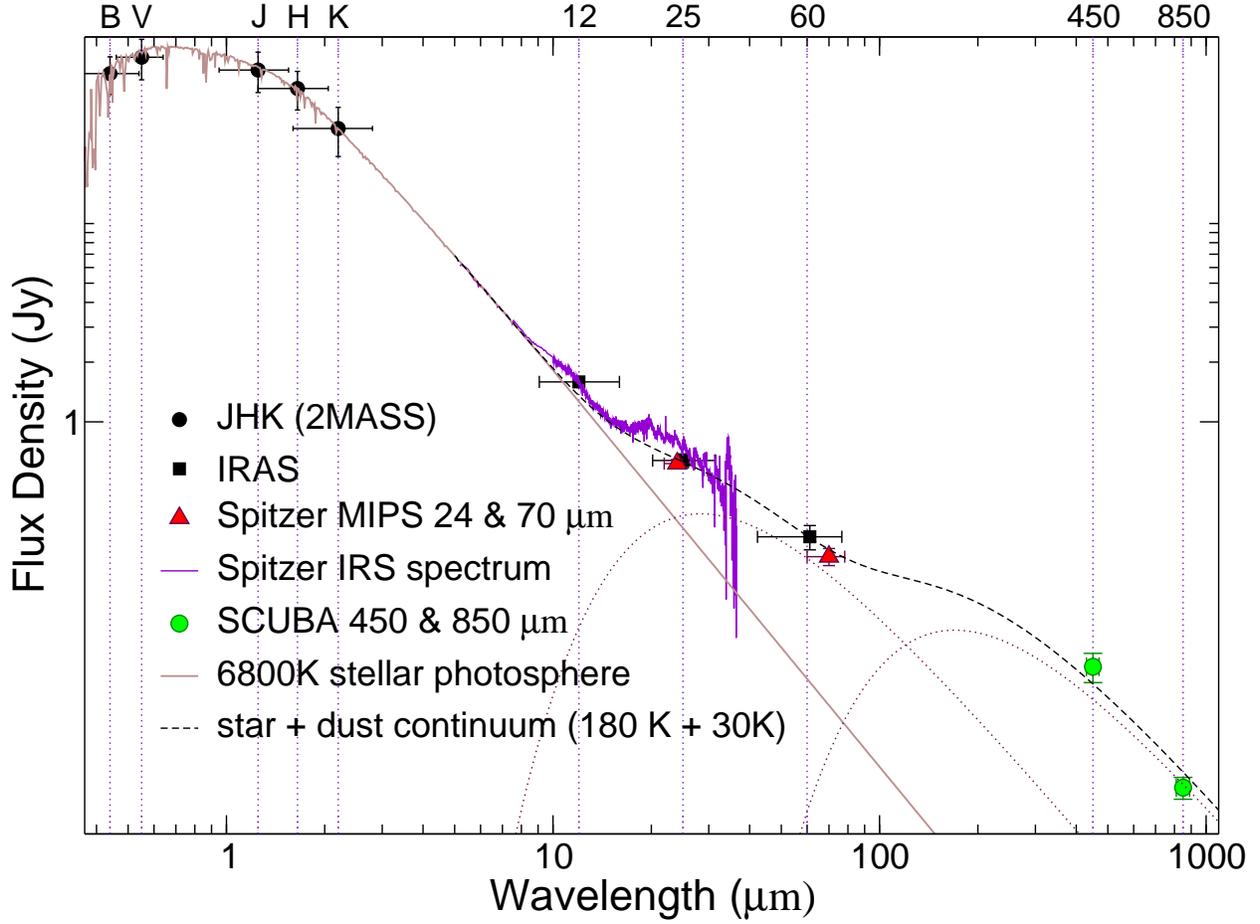}
\caption{Spectral energy distribution of $\eta$ Crv.  This should replace
the SED in Fig. 4 in \citet{wya05} because: (1) we plot {\it IRAS} 
FSC data in preference to the less accurate {\it IRAS} PSC data used by Wyatt 
et al (2005), and (2) we plot the Spitzer 24 micron MIPS measurement.  
Note that, as a result of these changes, our decomposition of the infrared
portion of the SED is very different from that of Wyatt et al.  Christine 
Chen kindly provided the flux-calibrated {\it Spitzer} IRS spectrum of 
$\eta$ Crv.}
\end{figure}

\end{document}